\documentclass[11pt]{article}
\usepackage{moriond,epsfig}

\def\Journal#1#2#3#4{{#1} {\bf #2}, #3 (#4)}

\def\be{\begin{equation}}
\def\ee{\end{equation}}
\def\bea{\begin{eqnarray}}
\def\eea{\end{eqnarray}}

\newcommand{\roots}{$\sqrt{s}$}
\newcommand{\invfb}{$\mathrm{fb}^{-1}$}
\newcommand{\invpb}{$\mathrm{pb}^{-1}$}
\newcommand{\pp}{$pp$}

\newcommand{\ttbar}{$t\overline{t}$}

\newcommand{\intlum}{$\int${\it L}dt} 
\newcommand{\pt}{$p_{T}$}
\newcommand{\et}{$E_{T}$}
\newcommand{\etmiss}{${E}_{T}^{miss}$}
\newcommand{\dilep}{$\ell\ell$}
\newcommand{\ljets}{$\ell$+jets}
\newcommand{\sigmattbar}{$\sigma_{t\overline{t}}$}
\newcommand{\sigmat}{$\sigma_{t}$}
\newcommand{\eb}{$\epsilon_{b}$}
\begin{document}
\vspace*{4cm}
\title{Top Physics at ATLAS and CMS}

\author{Francesco Span\`o \\(on behalf of the ATLAS and CMS collaborations)
}

\address{Columbia University, Nevis Laboratories, 136 South Broadway, P. O. Box 137,
\\ Irvington, NY 10533, USA}

\maketitle\abstracts{
%The Large Hadron Collider (LHC) will be a factory for top quark production.
The potential for top quark physics of the ATLAS and CMS experiments at the Large Hadron Collider is surveyed ranging from top quark ``re-discovery'' and its use as a calibration tool to initial and later stage measurements.}
\vspace{-0.9cm}
\section{Introduction: top quark at the LHC}

%\subsection{LHC and its multi-purpose experiments}
ATLAS~\cite{atlasdet} and CMS~\cite{cmstdrdet} are complementary
multi-purpose detectors aimed at measuring the properties of leptons,
hadrons and photons in proton-proton (\pp) collisions at the Large Hadron
Collider~\cite{mikelamont} (LHC).
On its way to reach the design center-of-mass energy (\roots\ = 14 TeV)
and instantaneous luminosity ({\it L} = $10^{34}$ $\mathrm{cm}^{-2}$
$\mathrm{s}^{-1}$), in 2009-2010 the LHC is expected to run
at~\roots\ = 10 TeV, delivering an integrated luminosity (\intlum) of
the order of 0.2~\invfb.~\cite{mikelamont}
All results reported in this paper are derived from simulated
collisions at~\roots\ = 14 TeV.

%\subsection{Top production and signatures}
%~\footnote{Gluon fusion processes dominate over~\qqbar\ annihilations.} 
At LHC with~\roots\ = 14 TeV  top quark pair (\ttbar) production cross
section~\cite{ttbarxsec} (\sigmattbar) is about 900 pb with a
theoretical uncertainty of order 10\% and a resulting rate of about 0.9
Hz already at $L$ = $10^{33}$ $\mathrm{cm}^{-2}$ $\mathrm{s}^{-1}$. 
%: 100 times larger
%than in~\ppbar\ collisions at~\roots\ = 2 TeV with a slightly smaller
%theoretical uncertainty (around 6\% to 4\%). 
%The resulting rate for~\roots\ =14 TeV is about 0.9 Hz already at $L$ =$10^{33}$
%$\mathrm{cm}^{2}$ $\mathrm{s}^{-1}$. 
The single top quark production~\cite{singletopxsec} is dominated by {\em t}-channel
diagrams that account for about 76\% of the total cross section
(\sigmat) of about 320 pb.
For LHC with \roots\ = 10 TeV, \sigmattbar\ drops~\cite{ttbarxsec} to $\approx$ 400
pb and, while \sigmat\ remains about three times smaller
than~\sigmattbar, %~\cite{ratiosingletottbarxsec}, 
the relevant $W$+n jets
background  cross section (with n$>$1) decreases by $\approx$ 23\%: this results into a somewhat worse signal to background ratio (S/B)
for \ttbar\ in comparison to ~\roots\ = 14 TeV. 
%~\cite{Wplusnjdecrease}
%for~\ttbar\ pairs at~\roots\ = 10 TeV. 
%The resulting signatures in the detectors are determined by the fact
%that

% FIXME MENTION THE CHAINS
%\ttbar \rightarrow WWbb \rightarrow \ell \nu b$
As the top quark decays to a $W$ boson and a $b$-quark
 with a branching fraction ($BR$) of almost 100\%, the~\ttbar\ final state features a $b$-jet pair associated
with high transverse momentum (\pt) light jets when each $W$ boson decays to
quarks (fully hadronic channel with $BR$ $\approx$ 44.4\%) and one
(\ljets\ channel with $BR$ $\approx$ 44.4\%) or two high~\pt\
lepton(s) (\dilep\ channel  with $BR$ $\approx$ 11.2\%)
and sizeable transverse missing energy (\etmiss) in the cases where one or two W
bosons decay to leptons respectively.
The final states of single top quark and~\ttbar\ can be
obtained from one another by swapping one $t \rightarrow Wb$ leg of
the~\ttbar\ decay with a $W$ boson ($Wt$ channel) or
one/two quarks ($s$ and $t$-channels), one of which is a $b$-quark.  
%As a consequence 
The two final states then have similar backgrounds (single
bosons ($W$, $Z$) plus jets, di-bosons and Quantum Chromodynamics (QCD)
multi-jet events) and they are background to each other.

%Single top quark and \ttbar\ final states have similar backgrounds (single
%bosons ($W$,$Z$) plus jets, di-bosons and QCD multi-jet events) and they
%are background to each other.~\footnote{The final states of single top quark and the~\ttbar\ pair can be
%obtained from one another by swapping one $t \rightarrow Wb$ leg of
%the~\ttbar\ decay with a $W$ boson ($Wt$ channel) or
%one/two quarks ($s$ and $t$-channels), one of which is a $b$-quark.  }

\section{Re-discovering Top}
A clean and robust analysis is required to re-establish the top quark signal with early data. 

\begin{figure}[h]
\begin{center}
\psfig{figure=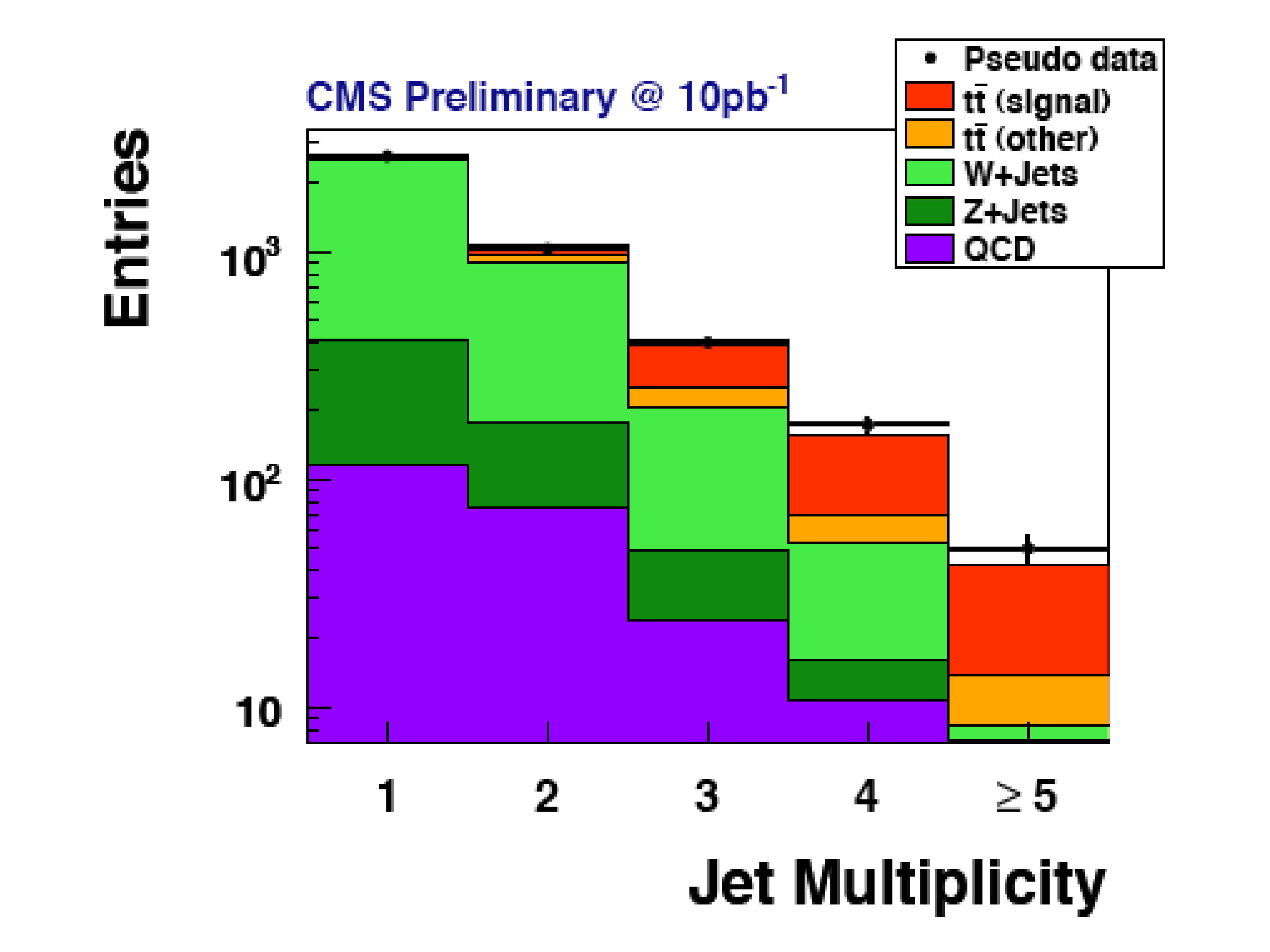,height=1.5in}
\psfig{figure=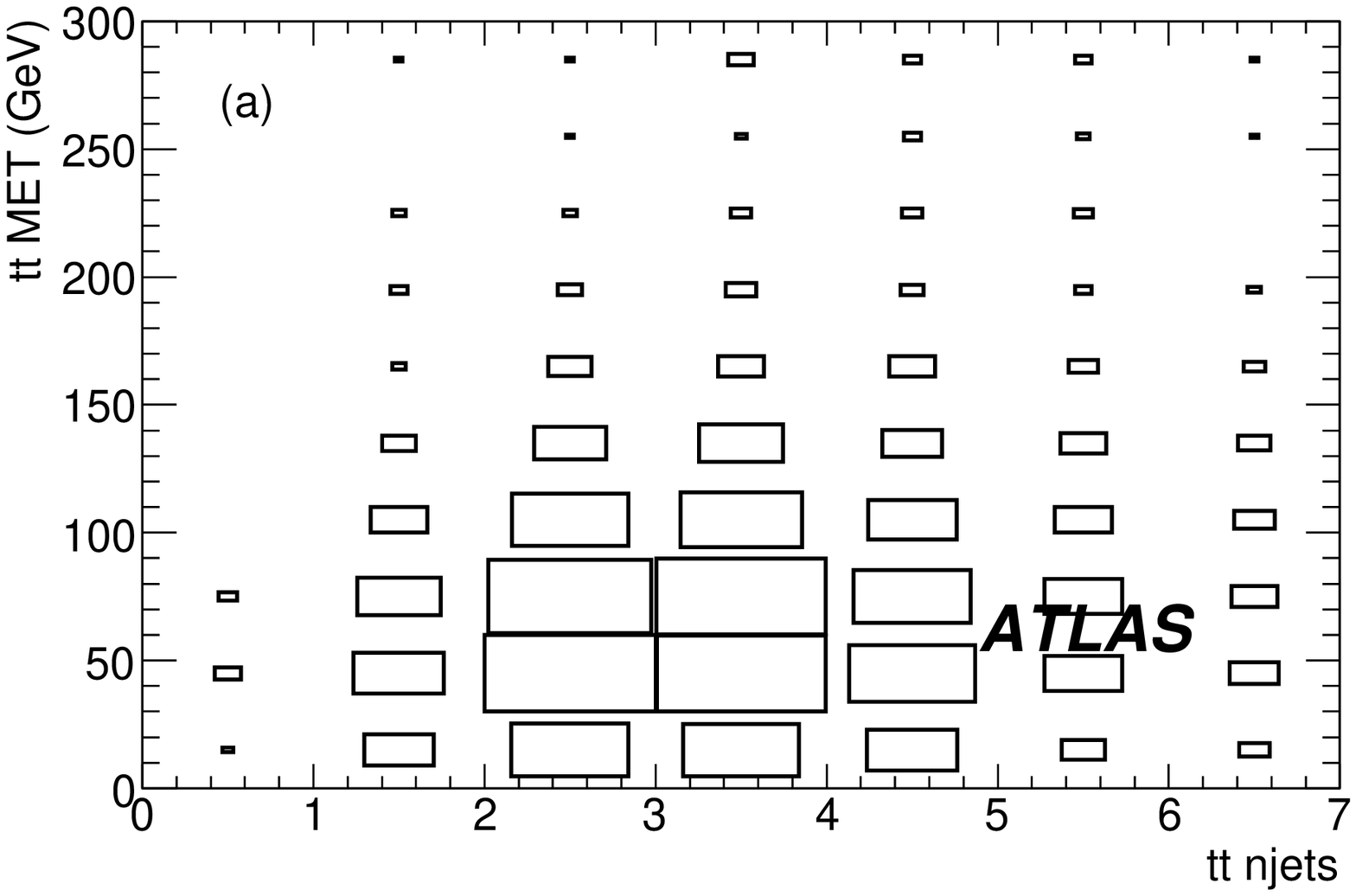,height=1.35in}
\psfig{figure=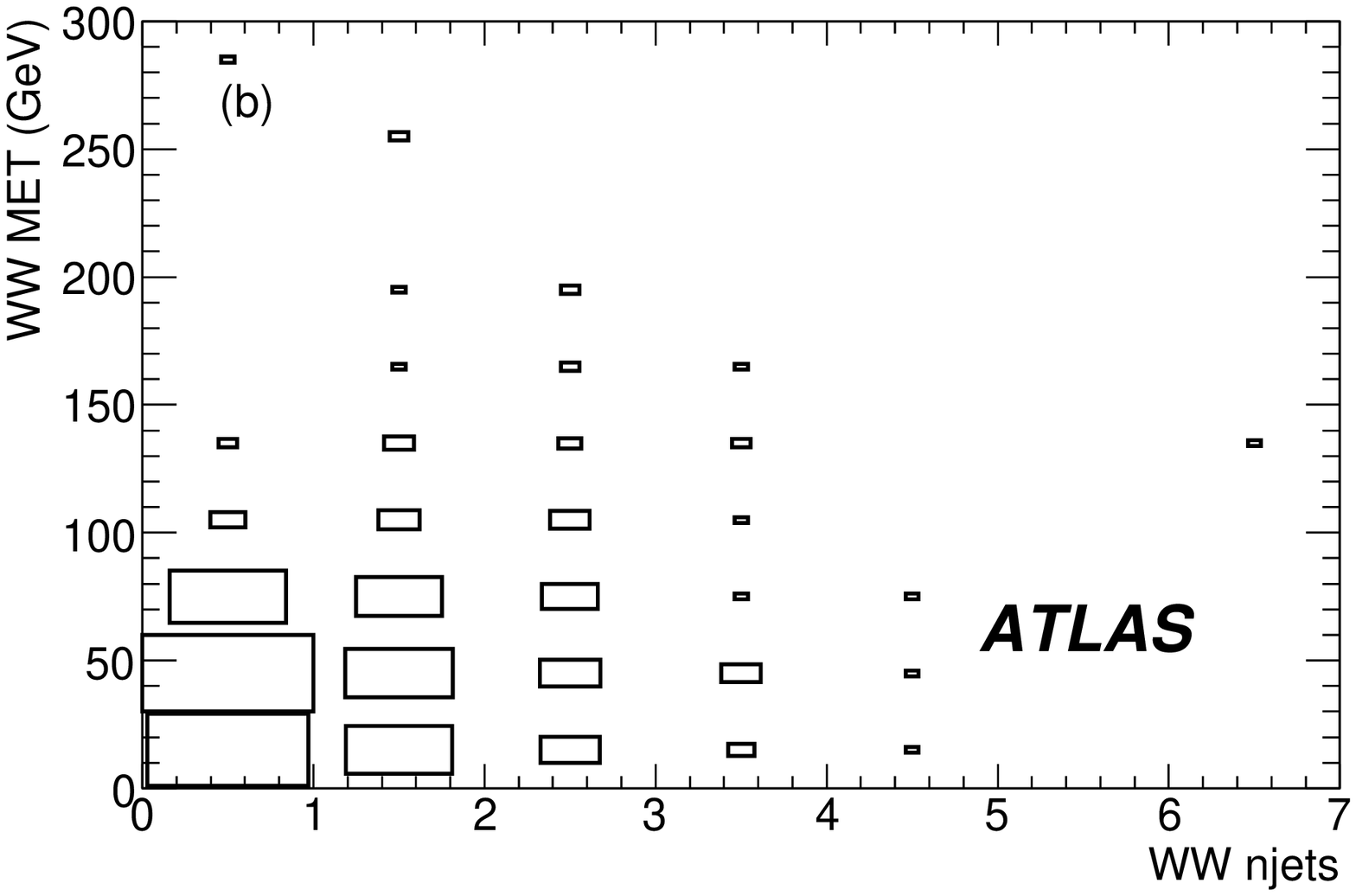,height=1.35in}
\end{center}
\caption{Left: CMS reconstructed jet multiplicity distribution for~\ttbar\ events passing the
  final selection with CMS detector except for the requirement $\mathrm{N}_{\mathrm{jets}}$
  $\geq$ 4. Center and Right: simulated $e\mu$ templates in the
  plane of \etmiss and number of jets for~\ttbar\ and $W$+jets
  samples as seen by the ATLAS
  detector. See text for references.\label{fig:rediscoveryanddilepxsec}}
\end{figure}

An example CMS analysis~\cite{cmsrediscover} uses a realistic
first day simulation of the detector mis-calibration and misalignment.
 The distinguishing features of the \ljets\ final state
 (one central high~\pt\ muon ($\mu$) and four or more central high~\pt\
jets) are coupled to additional $\mu$ isolation cuts (on energy deposited in a calorimeter
cone around the $\mu$ and the $\mu$ spatial distance from the closest jet) to achieve
a drastic reduction of the large QCD background.
% The addition
%isolation cuts involve the amount of energy deposited in a calorimeter
%cone around the muon and the space distance of the muon from the closest jet.
With only~\intlum\ = 10~\invpb\ the~\ttbar\ signal is expected to
emerge from the background in the high jet multiplicity bins
($N_{jets}$$\geq$ 4) with about 130 signal events and S/B $\approx$
1.4 as shown in figure~\ref{fig:rediscoveryanddilepxsec} (left). No $b$-tagging
or~\etmiss\ cuts are used. The size and shape of the QCD background
have large uncertainties and its data-driven determination is a
crucial ingredient in the analysis.
% ATLAS section 3.3.2 of Top quark 

Once the signal is established, the lower $BR$ \dilep\ final
state provides a clean sample to measure
\sigmattbar.~\cite{atlasxsec,cmsdilepxsec}
%~\footnote{{\em Notice that in
%  \ppbar\ collisions at~\roots = 1.9 TeV, CDF and D0 measure the cross
%  section in this channel with precisions ranging from 8.6\% to 36\%
%  for CDF (2.8 to 2 $\mathrm{fb}^{-1}$) and 11\% to 12\% for D0
%  respectively~\cite{dilepxsectevatron} using \intlum\ = 2.8\invfb\
%  (CDF) and \intlum\ = 1 \invfb\ (D0).}}. 
%%%%%(from the two neutrinos in the final state)
The selections of both ATLAS and CMS require two high~\pt\
isolated leptons with opposite charges and large \etmiss\
 that is not collinear with any of the two leptons or with the leptonic system.
In an example from ATLAS \sigmattbar\ is derived even
without any additional cuts (no $b$-tagging) by a likelihood fit of the
signal to simulated templates of distributions in the~\etmiss-jet multiplicity
space.~\footnote{Other methods use additional cuts on the number of
  jets and their transverse momentum to purify the sample. With no
  $b$-tagging the cross section is extracted by counting the events in
  multiplicity bins or performing a likelihood fit to angular
  variables (ATLAS)~\cite{atlasxsec} while the  use of $b$-tagging and W
  mass constraint can also be added (CMS).~\cite{cmsdilepxsec}}
As one can see in figure~\ref{fig:rediscoveryanddilepxsec} (center and
right), \ttbar\ events show higher multiplicity and somewhat higher
\etmiss\ than $W$+jets background.
With \intlum\ = 100~\invpb\ such an analysis~\cite{atlasxsec,cmsdilepxsec} has an outstanding
significance ($\approx$ 20) which is already
$\approx$ 7 with \intlum\ = 10~\invpb: the \dilep\ channel can also be
used to establish the \ttbar\ signal.
%~\cite{cmsdilept10pb}.
%~\footnote{a small
%  statistical  uncertainty is already achievable with \intlum=10~\invpb }.
For \intlum\ = 100~\invpb\ the ATLAS fractional uncertainty on \sigmattbar\
($\delta$\sigmattbar/\sigmattbar) shows comparable systematic and statistical
contributions (7\% and 4\% respectively).~\footnote{CMS estimates that
  for \intlum\ = 1 \invfb\ the statistical contribution is expected to drop to 0.9\%
  compared to a dominant systematic contribution of $\approx$
  11\% (see section 8.1.2 of ~\cite{cmstdrphys}).} 
The crucial analysis requirements are to validate
the~\etmiss\ description and perform data-driven background estimates.

%\begin{figure}
%%\psfig{figure= T6_fig15_right_significance.eps,height=1.35in}
%\psfig{figure=Topmass100pb_selB_colour.eps,height=1.5in}
%\captionsr{ Left: Projected significance from pseudo-experiments
%  for the template fit as a function of integrated luminosity
%  including all channels and all systematic uncertainties.
%Right:Reconstructed top mass after W boson mass constraint for electron analysis.
%\label{fig:dilepxsecatlas}}
%\end{figure}
\begin{figure}
\psfig{figure=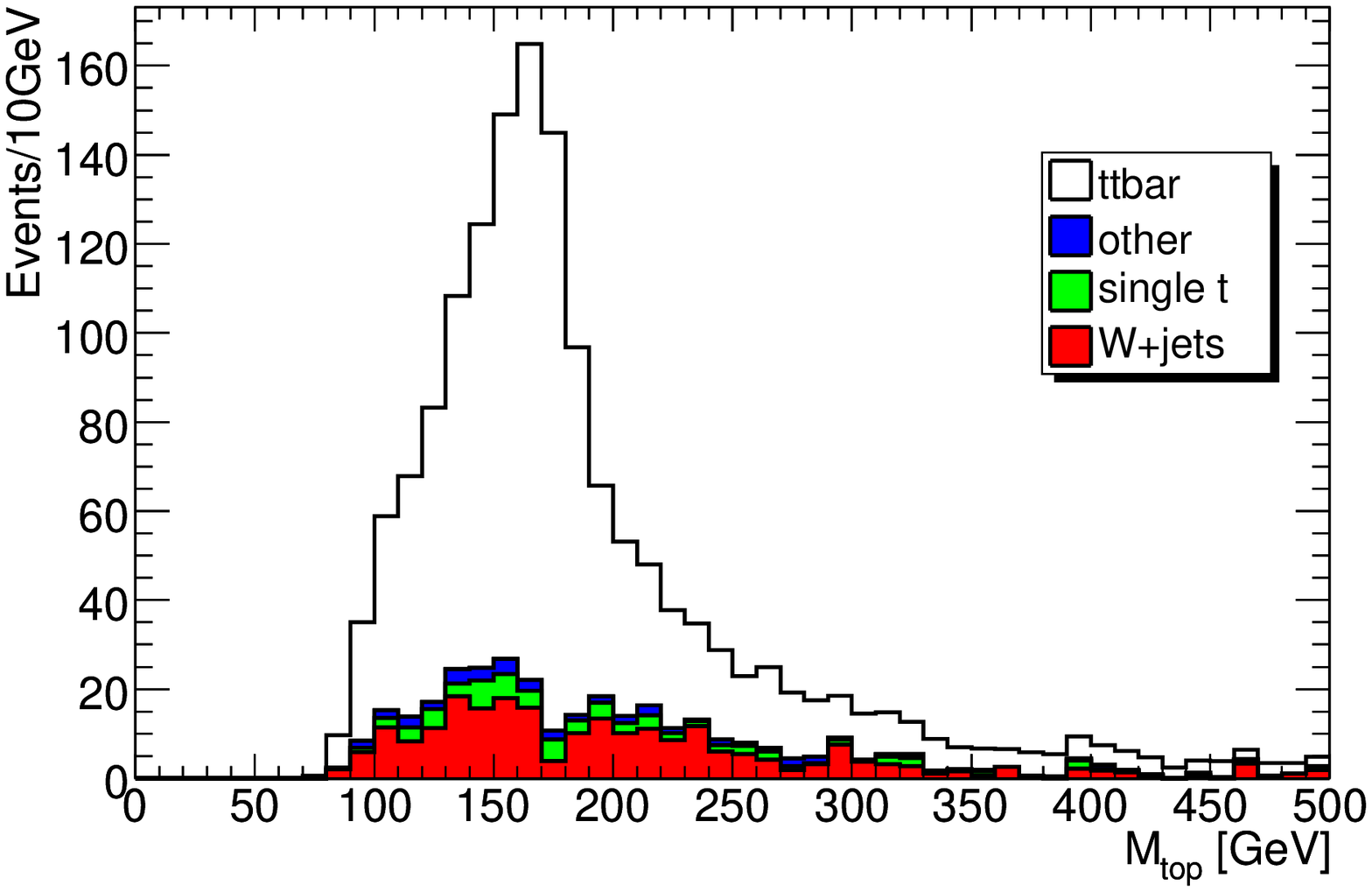,height=1.5in}
\psfig{figure=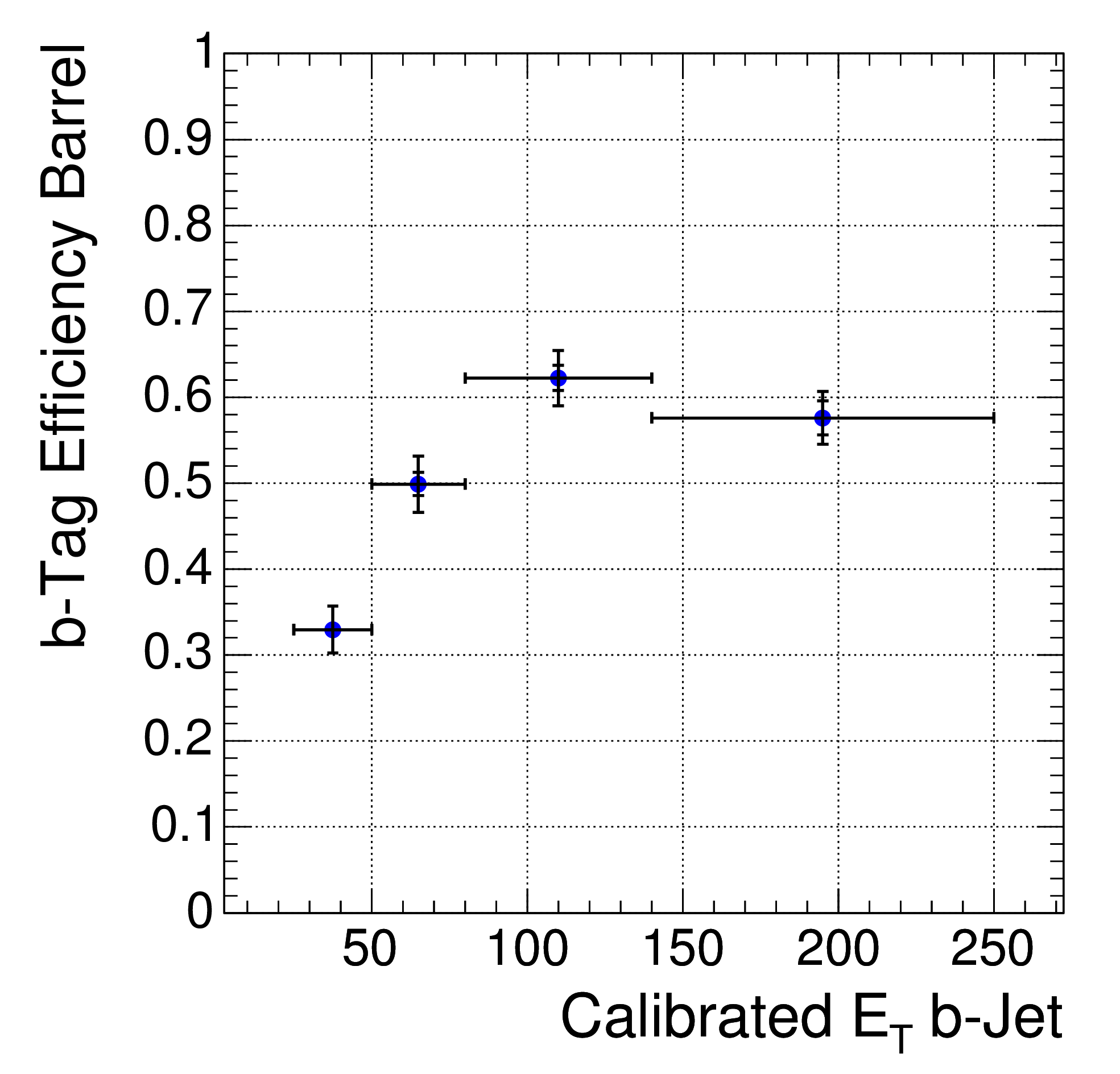,height=1.5in}
\psfig{figure=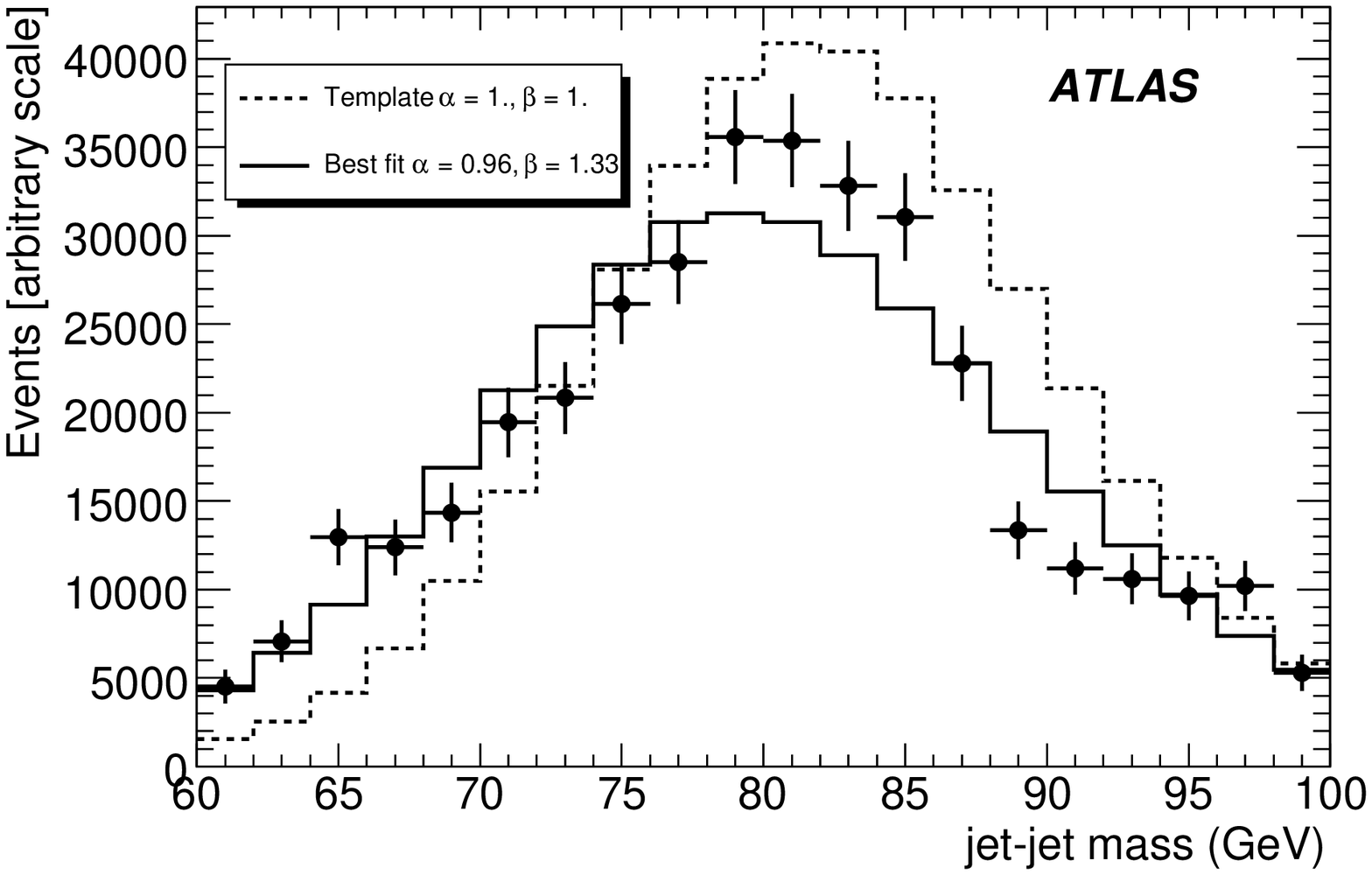,height=1.5in}
\caption{Left: ATLAS reconstructed top quark mass after $W$ boson mass constraint for semi-leptonic ($\ell = e$) analysis.
Center: CMS $b$-tagging efficiency measurement as a function of the jet
  $E_{T}$ for jets in the barrel ($|\eta|$ $<$ 1.5).
  Right: ATLAS di-jet invariant mass in fully simulated \ttbar\
  events (dots) superposed on the template histogram with $\alpha$ = 1
  and $\beta$ = 1 and the best fit histogram. See text for references.\label{fig:xsecandcalib}}
\end{figure}

%~\footnote{Notice
% s that in~\ppbar collisions \roots\ = 1.9 TeV, CDF  
% and D0 measure the total cros section in with uncertainties
%  around 14\% to 18\% respectively~\cite{dilepxsectevatron} using
%  \intlum\ = 2.8 $\mathrm{fb}^{-1}$ (CDF) and \intlum\ = 1
%  $\mathrm{fb}^{-1}$ (D0)} 
The more statistically powerful \ljets\ channel ($\ell$ = $e$,$\mu$) is also
used to extract \sigmattbar\ by requiring one high~\pt\ central lepton and
four or more central high~\pt\ jets. The reconstruction of the hadronic
top is used to enhance the signal over the dominant $W$+jets background
when no $b$-tagging is used. Figure~\ref{fig:xsecandcalib} (left) shows
the ATLAS result~\cite{atlasxsec} for \intlum\ = 100~\invpb\
where the hadronic top consists of the three jets with the highest
total \pt\ and at least one di-jet pair is required to have a mass consistent with
the W boson.~\footnote{The addition of one or two $b$-tagged jets
  improves S/B~\cite{atlasxsec}
  (ATLAS) and with two $b$-tagged jets and cleanly separated jets, a
  convergent kinematic fit imposing the W mass constraint to the light
  di-jet system can be used to select events (see section 8.1.3 of
  ~\cite{cmstdrphys}) (CMS).}
 Then \sigmattbar\ can be extracted by a likelihood fit to the mass
 shape (ATLAS) or by simply counting events after subtracting the
 expected background (ATLAS, CMS).
 For \intlum\ = 100~\invpb\ the ATLAS $\delta$\sigmattbar/\sigmattbar\ 
 in a robust counting experiment~\cite{atlasxsec} is already systematic-dominated
 ($\approx$ 17\% compared to 3\% statistical contribution).
The main initial systematic uncertainties are the jet energy scale
 (JES) and the normalization of the W +jets background which, like the
 estimate of the uncertain QCD background, benefits from a data driven
 approach. A precise (5\%) measurement of the b-tagging efficiency will be 
necessary for it to be used  effectively in the cross-section measurement.

\section{Top for Calibration}

The established \ljets\ sample can be used (together with the
di-lepton (CMS)) to measure the $b$-tagging efficiency \eb. By
exploiting cuts on the kinematic and topological properties or by
cutting on likelihood discriminants, it is possible to select a highly
enriched b-jet sample in which, after background
subtraction, $\epsilon_\mathrm{b}$ can be measured~\cite{atlasbtag}, also as a
function of the $b$-jet pseudo-rapidity and transverse energy
(\et)(also see section 12.2.8.1 of~\cite{cmstdrdet}).
%{cmsbtag} 
CMS shows an example of this determination  in
figure~\ref{fig:xsecandcalib} where \eb\ is
shown to vary from 30\% to 60\%  in the \et\ range
(50 GeV, 250 GeV) for the barrel section using
\intlum\ = 1\invfb. With \intlum\ = 1~\invfb\ an uncertainty on
$\epsilon_\mathrm{b}$ of 6\% in the barrel  and of 10\% in the endcaps
is expected.\footnote{A 10\% uncertainty is expected by
  ATLAS for \intlum\ = 100~\invpb.}
 ATLAS showed that a global average \eb\ can be known at the 5\% level
 with \intlum\ = 100~\invpb by performing a
likelihood fit to the expected number of events with zero to three
$b$-tagged jets~\cite{atlasbtag} (\eb, the $c$-tagging efficiency and \sigmattbar\
 are determined simultaneously).
% For  d\eb/\eb \approx 5\%.
%the resulting uncertainty on the global
%$\epsilon_\mathrm{b}$ ranges from 5\% to 10\% with different
%techniques (see~\cite{btagdiffatlas}).
% using
%100~\invpb of integrated luminosity.
%\begin{figure}
%\rule{5cm}{0.2mm}\hfill\rule{5cm}{0.2mm}
%\vskip 2.5cm
%\rule{5cm}{0.2mm}\hfill\rule{5cm}{0.2mm}
%\psfig{figure=Figure_012-036-a_CMSbtag.ps,height=1.5in}
%\psfig{figure=T2_fig21_dijetmasstemplate.eps,height=1.5in}
%(see section 12.2 of ~\cite{btagcms}).
%(see section 12.2 of ~\cite{btagcms}).
%\caption{b-tagging egfficiency measurement, as a function of the jet
%  $\mathrm{E}_{\mathrm{T}}$ for jets in the barrel ($|\eta|$ $<$ 1.5)
%  \label{fig:effbtagcms} Di-jet invariant mass in fully simulated
%  events (dots) superposed on the template histogram with $\alpha$=1
%  and $\beta$ = 1 and the best fit histogram  b-tagging efficiency
%  measurement, as a function of the jet $\mathrm{E}_{\mathrm{T}}$ for
%  jets in the barrel ($|\eta|$ $<$ 1.5). \label{fig:jetenscaletemplateatlas}}
%\end{figure}

Once $b$-tagging is understood the \ljets\ sample can be used to identify
the $b$ and the $W$ di-jet system that result from the ``hadronic''
top decay (for high \pt\ jets). In this way in ATLAS the global average JES for
light jets is expected to be known at the 2\% level even with 50 \invpb\ by
performing a $\chi^{2}$ fit to di-jet mass simulated templates when
varying the overall light jet scale ($\alpha$) and resolution ($\beta$)
with respect to the default.~\cite{atlasbtag} 
An example of such technique is shown in
figure~\ref{fig:xsecandcalib} (right).
Differential information on the light JES can be known
at about 2\% with \intlum\ = 1~\invfb\ by using re-scaling
techniques~\cite{atlasbtag}, while CMS expects that exploiting the
top quark mass measured at the Tevatron can provide knowledge of the average $b$
JES~\cite{jetenscalecms} at the \% level with \intlum\ = 100~\invpb.

\section{Measuring Top}

\subsection{Top Quark Mass}

\begin{figure}
%\rule{5cm}{0.2mm}\hfill\rule{5cm}{0.2mm}
%\vskip 2.5cm
%\rule{5cm}{0.2mm}\hfill\rule{5cm}{0.2mm}
\begin{center}
\psfig{figure=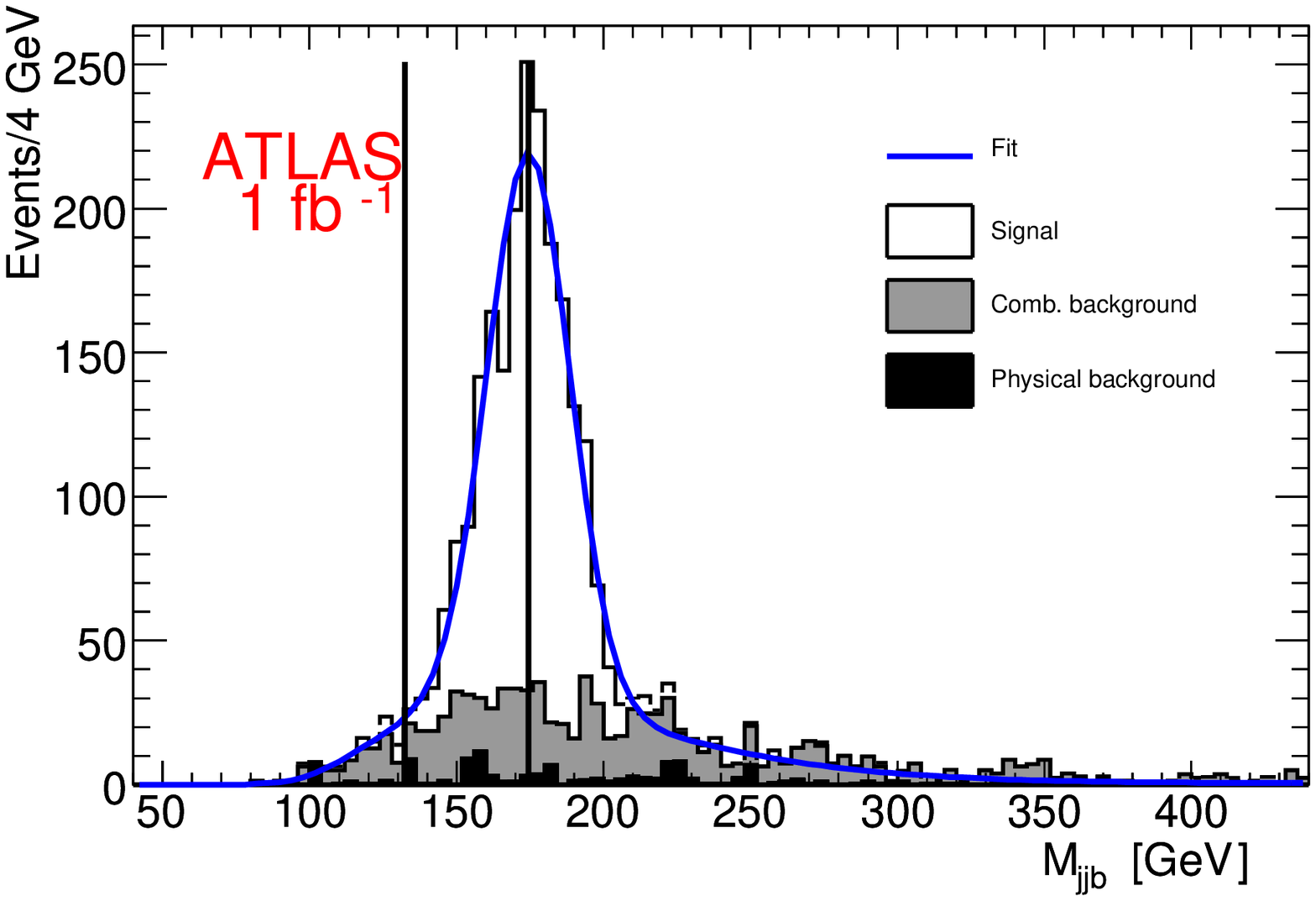,height=1.5in}
\psfig{figure=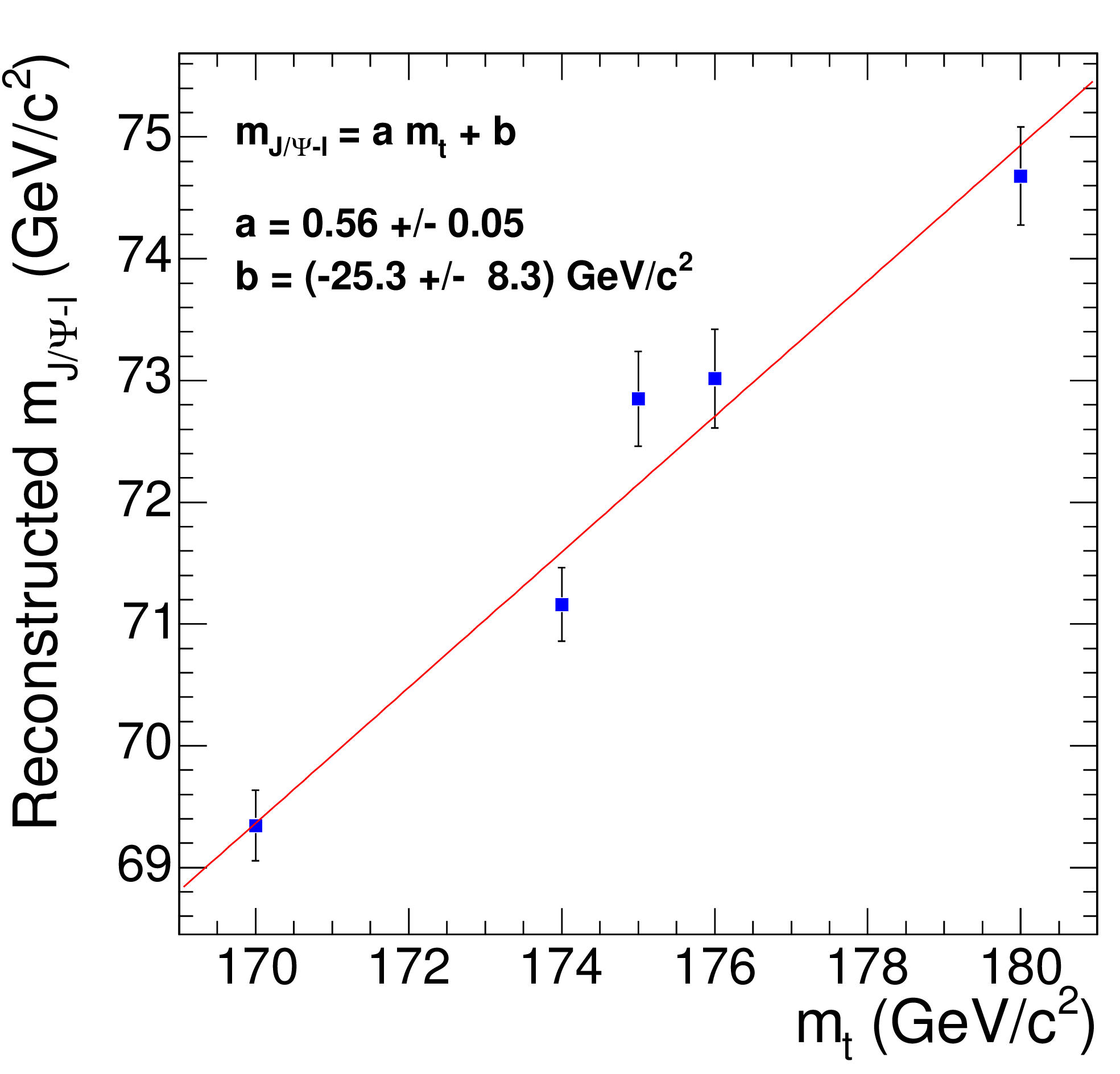,height=1.5in}
\end{center}
\caption{Left: ATLAS reconstructed hadronic top quark mass fitted with the sum of a Gaussian
  and a ``threshold'' function, scaled to \intlum = 1 \invfb. Right: correlation
  between the CMS reconstructed three lepton invariant mass and the top
  input mass at full simulation. See text for references.}
% left (section 3.8) of~\cite{atlascsc})  
% right(see section 8.2.4.3 of~\cite{cmstdrphys}}
\label{fig:topmassatlascms}
\end{figure}

Equipped with better understanding of $b$-tagging and JES,
the top quark mass ($m_{top}$) can be measured in the \ljets\ channel by using the
same selection required for \sigmattbar\ measurements and
requesting at most harder jets (ATLAS) to suppress
backgrounds. The hadronic $W$ is reconstructed even with
minimal or no $b$-tagging information and it is then associated with the
closest jet or $b$-jet by its spatial (ATLAS) or kinematic (ATLAS/CMS)
distance.~\cite{topmassatlas,topmasscms} 
% reconstructed with and without the use of b-tagging The hadronic W
% is  a selection and b-tagging, a simple geometry based
% reconstruction is coupled to cuts on kinematic variables to
% associate one of the tagged b-jets to the hadronic
% W~\cite{atlastopmasswdetail}. 
The top quark mass value is then derived by either fitting an analytic
function (ATLAS), extrapolating  or performing a more sophisticated
event-by-event likelihood fit (CMS).
An example from ATLAS is shown in figure~\ref{fig:topmassatlascms}:
after $b$-tagging, a nearly background-free scenario is obtained where
the reconstructed top mass is fitted by a Gaussian plus a ``threshold
function''.~\cite{topmassatlas}
The measurement is quickly systematics dominated, mainly by the jet
energy scale (particularly the $b$-jet scale).
A top quark mass uncertainty of the order of 1 to 5 GeV is expected to be
achievable~\cite{topmassatlas,topmasscms} with \intlum\ = 1~\invfb\ if
the uncertainty on the JES is in the range of 1\% to 5\%
respectively.
%~\footnote{The current  measurement from the
%  Tevatron~\cite{mtoptev} gives $m_{top}$(CDF+D0)
%  = 173.1 $\pm$ 1.2 GeV (up to 3.6 \invfb)}.
The fully hadronic and \dilep\ channel also have mass
information, but the extraction is harder due to the increased level of combinatoric
background and the final state
neutrinos.~\footnote{For instance CMS expects~\cite{topmasscms} top
  quark mass uncertainties
  of about 1.2 GeV in the \dilep\ channel with \intlum\ = 10~\invfb\ and
  4.2 GeV in the fully hadronic channel with \intlum\ = 1 \invfb.~\cite{topmasscms}}

%It is then nteresting to consider an alternative technique to
%measure the top mass. 
Alternative techniques to measure the top quark mass are also
considered to reduce the impact of JES systematic uncertainty.
In an example from CMS~\cite{topmassfromjpsicms},
  \ljets\ events are sifted through to find exclusive $b$-jet decays
  to J/$\Psi$ (with the b-jet coming from $t \rightarrow Wb\rightarrow
  \ell \nu b$ chain).
%In an example from CMS~\cite{topmassfromjpsicms}, after selecting
%  \ljets\ events, a search is performed for rare J/$\Psi$ production in the
%exclusive decay of the $b$-jet (BR \approx $7 \cdot
%  10^{-4}$~\cite{pdg}) from the $t \rightarrow Wb\rightarrow \ell \nu
%  b$ chain  is searched for. 
The top quark mass is strongly correlated with
the mass of the system formed by the J/$\Psi$ and the lepton from the
``leptonic'' top decay as it is shown in the calibration curve by CMS in
figure~\ref{fig:topmassatlascms}.
The systematic uncertainty is dominated by theoretical contributions
(mainly from models of b-quark fragmentation and underlying event)
while the JES contribution is negligible. Given the low BR rate for
\ttbar\ event to produce a final state with a leptonic
J/$\Psi$~\cite{topmassfromjpsicms}, a top quark mass uncertainty of
about 2 GeV is expected for \intlum = 20~\invfb.

\subsection{Single top quark}
\begin{figure}
\begin{center}
\psfig{figure=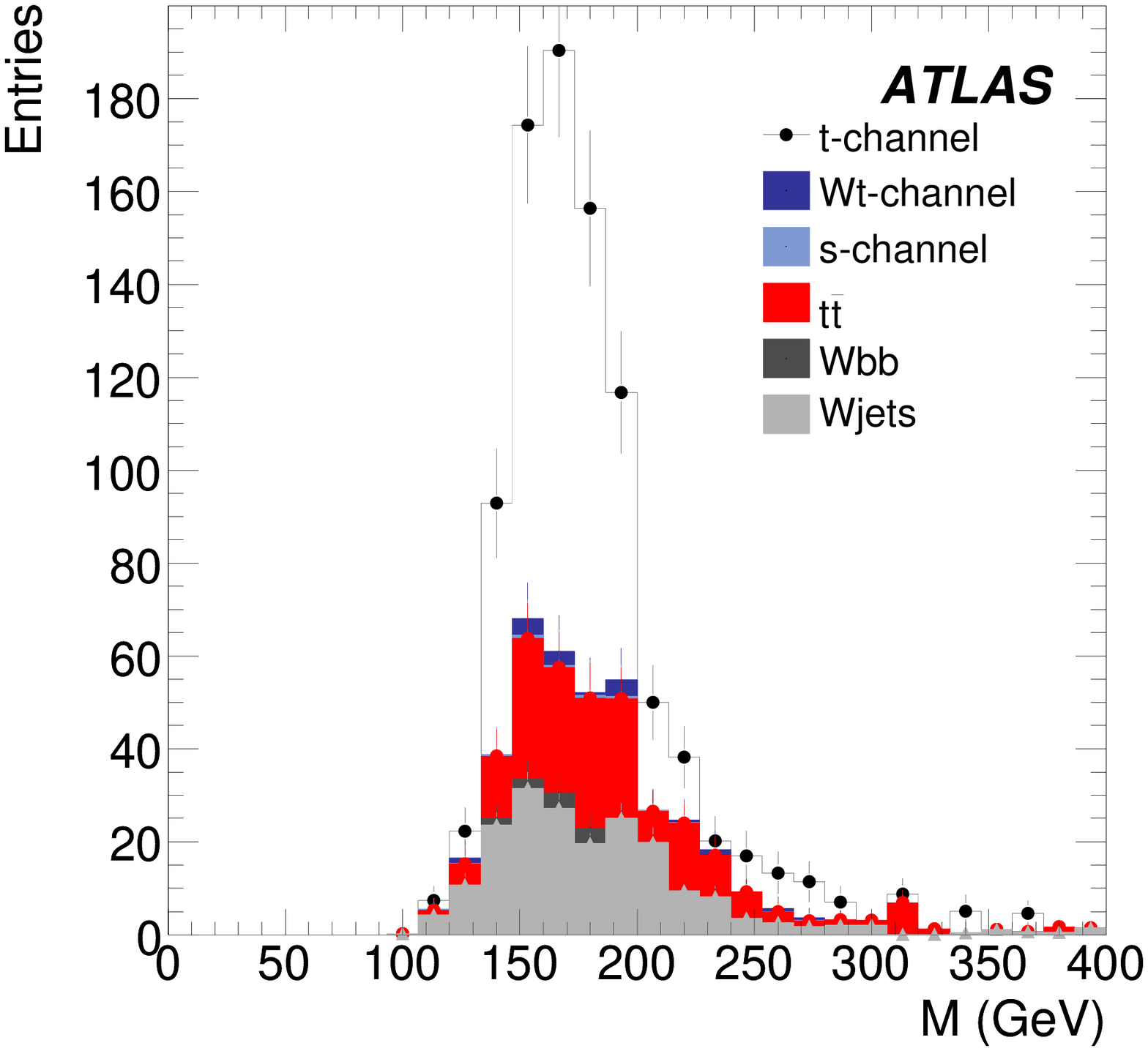,height=1.5in}
\psfig{figure=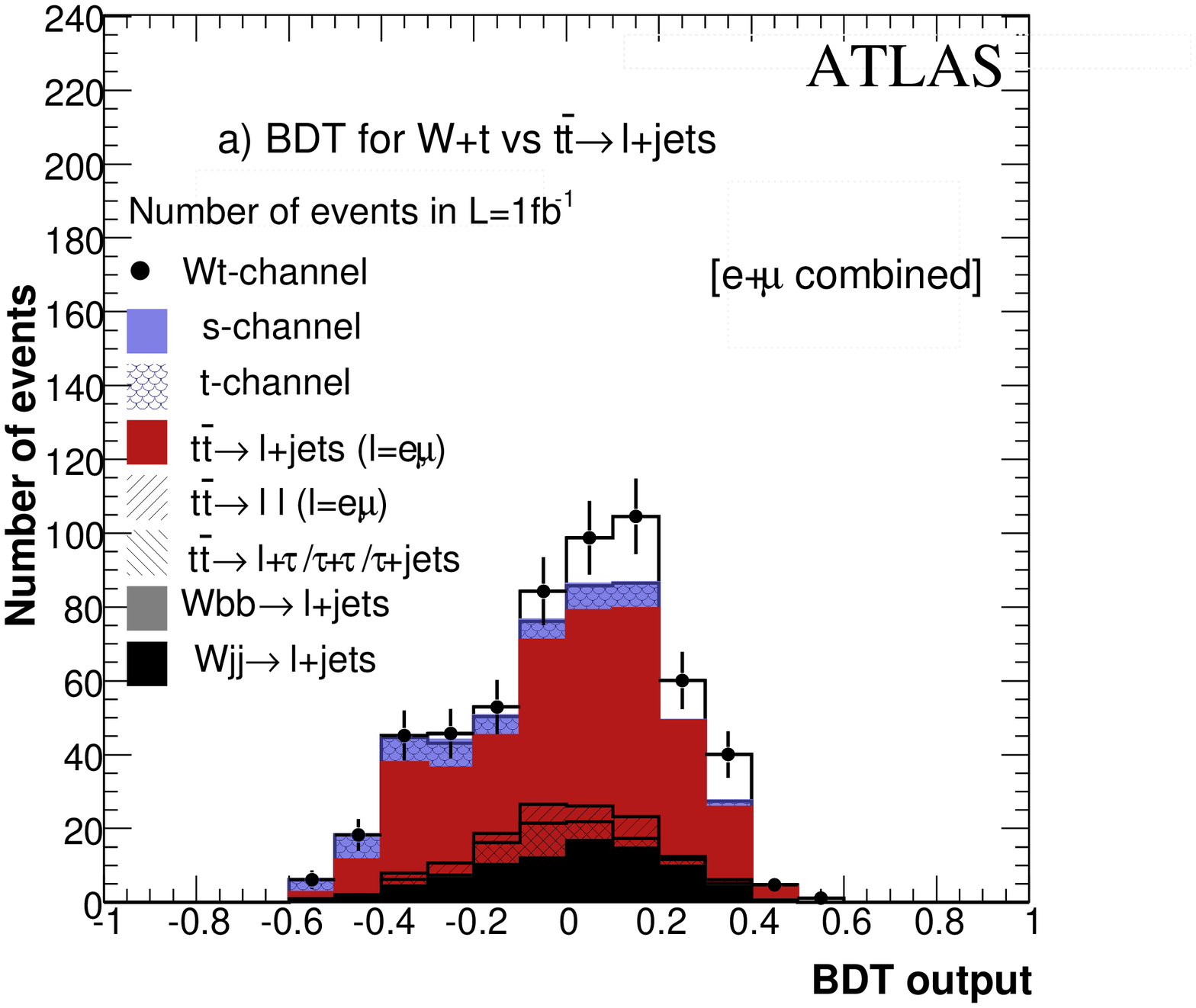,height=1.5in}
\psfig{figure=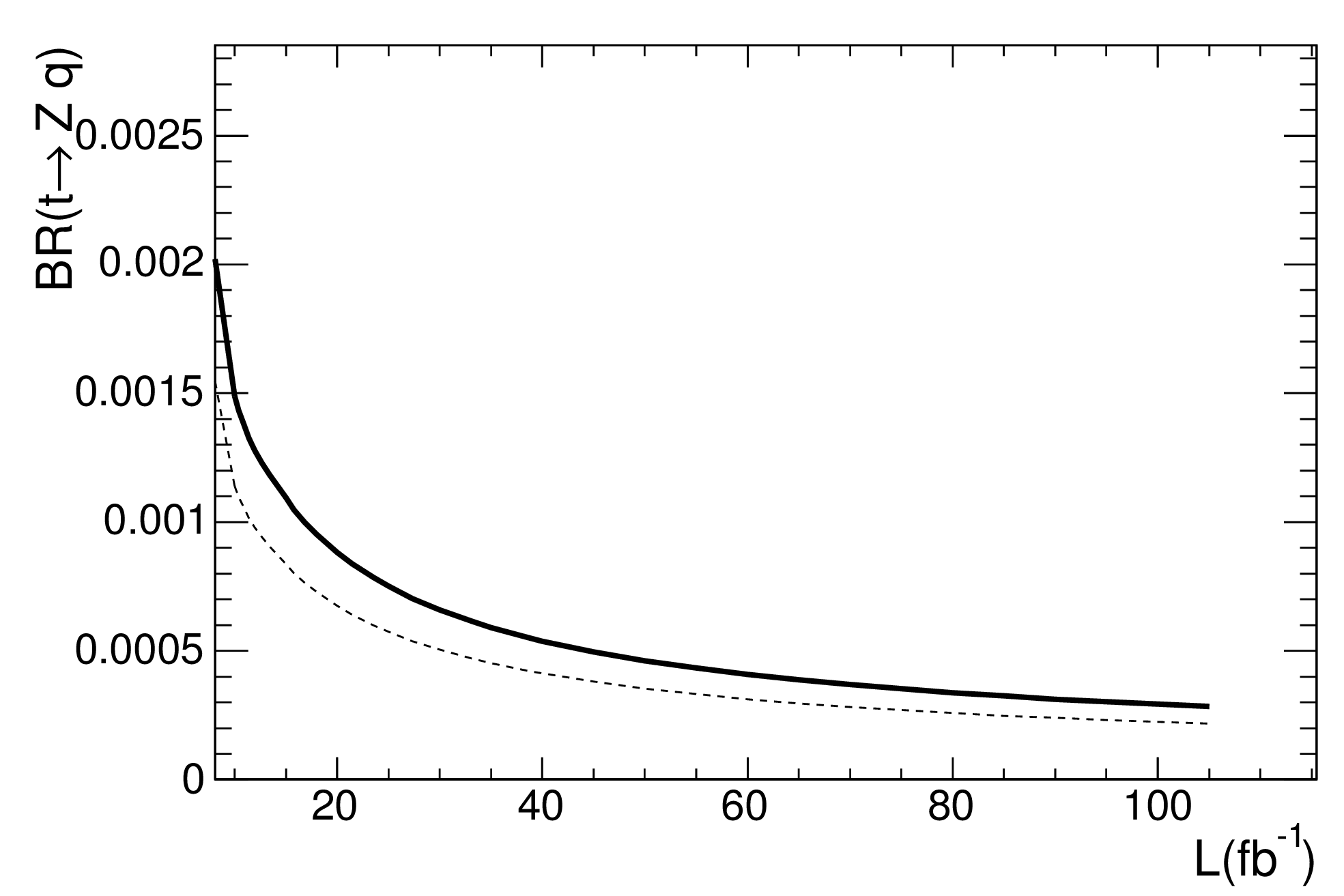,height=1.5in}
\end{center}
%Left:(see section 4.3 of~\cite{atlascsc})
%Center:(see section 6.2.of ~\cite{atlascsc})
%Rght:(see section 8.5.4 of~\cite{cmstdrphy})
\caption{Left: ATLAS reconstructed leptonic top quark mass distribution with cut on BDT
  output of 0.6. Center: ATLAS reconstructed distribution for BDT
  against~\ttbar in the~\ljets\ channel in the 3 jets final state analysis
 for~\intlum\ = 1\invfb. Right: branching ratios of an FCNC signal
  detectable by CMS at the 5 s.d. level as a function of the integrated luminosity
  for the $qZ$ channel, shown with (solid line) and without(dashed line)
  systematic uncertainties. See text for references.
%{ the five  standard deviation}\
 %BDT output defined in the 3  jet ('lb2j')
%  final state analysis }
\label{fig:singletopandfcnc}}
\end{figure}

Once the~\ttbar\ signal is established and measured, the attention
can turn to measuring \sigmat\ in the dominant
$t$-channel where one top quark decays leptonically (the final state
is then $q \ell \nu b$($b$)).
In an example from ATLAS,~\cite{singletopatlas} a complex set of
kinematic cuts (on the final state lepton, the $b$-jet and the
distinctive single forward light jet) is coupled to a multivariate
Boosted Decision Tree (BDT) based on a set of shape variables to
separate the signal from the background. The distribution for the top
quark mass after all cuts for~\intlum\ = 1~\invfb\ is shown in
figure~\ref{fig:singletopandfcnc} (left) where the single top signal is
standing out of the remaining~\ttbar\ and $W$+jets backgrounds.
With these techniques S/B is expected to
be about 1.3 with $\delta$\sigmat/\sigmat\ 
%a percentage uncertainty on the cross section 
dominated by systematic contributions even for~\intlum\ = 1~\invfb.
Given the sizeable background to be overcome, the~\ttbar\ signal
knowledge from the data is required in addition to the $W$+jets and the
QCD signals. In addition an excellent detector understanding is required
($b$-tagging performance, JES) also to control the BDT inputs.

The less statistically powerful $Wt$ channel is very similar to the~\ttbar\ signal in its final state 
(one $b$-jet less). So, as ATLAS
shows in figure~\ref{fig:singletopandfcnc} (center), even after cuts on
a set of kinematic variables and a set of BDT outputs based on
twenty-five variables, the~\ttbar\ background is still
sizeable~\cite{singletopatlas}. This results into an expected S/B of about 0.4
with \intlum\ = 1 \invfb. Only a few~\invfb\ will allow three standard
deviation (s.d.) evidence to be established and a measurement of the
cross section with a relative uncertainty of 20\% should be in sight
with 10 \invfb. The lowest BR $s$-channel is expected to be the most
difficult: ATLAS, for instance, expects to achieve a 3
s.d. evidence~\cite{singletopatlas}
%standard deviations%
with about 30~\invfb.~\footnote{CMS expects to need at least
10~\invfb\ to measure the cross section with a systematic-dominated
uncertainty of about 30\%.~\cite{schancms}}  
Even more than for the $t$ channel, both $s$ and $Wt$ channel will require
an excellent detector understanding to be coupled to data-driven
background estimation techniques and to a good control of the
theoretical description of initial and final state radiation
(particularly for the $Wt$ channel whose similarity to~\ttbar\ makes
it more sensitive to jet multiplicity).

\section{Top beyond the standard model}
Measurements of the properties of the top quark offers a window onto
possible physics beyond the standard model.
%Top properties are windows on possible physics beyond the standard model.
An example from CMS~\cite{fcnccms} shows the sensitivity to Flavour
Changing Neutral Current events where a $t \rightarrow Wb$ leg of the~\ttbar\ decay
is replaced by a $t \rightarrow Zq$ decay. The resulting final state
is $\ell \ell q l \nu b$. 
Cuts are applied to all the expected final state particles: one $b$-tagged jet,
one light jet and two isolated opposite-sign leptons whose di-lepton mass
is consistent with the $Z$ boson. Additional cuts on the masses and
the angular relation of the $Zq$ and $Wb$ systems tend to select
events consistent with a~\ttbar\ decay. Once the background is
subtracted (mainly \dilep~\ttbar\ events) it is possible to count the
events and measure the cross
section. Figure~\ref{fig:singletopandfcnc} (right) shows the cross sections for which a
5 s.d. sensitivity is expected as a function of the
collected integrated luminosity. With \intlum\ = 10~\invfb\  a
branching ratio of 1.49 $\cdot$ $10^{-3}$ is expected
to be detectable (see section 8.5.4
of~\cite{cmstdrphys}). The main systematic uncertainties derive from jet
and lepton energy scale and $b$-tagging efficiency.
%five standard deviations  
%\section{Conclusions}
%The LHC will be a top factory (both at \roots = 10 TeV and \roots = 14
%TeV) where most measurements will become systematics dominated quickly.

%At \roots = 14TeV, ATLAS and CMS can accomplish:
%\begin{itemize} 
%\item Top re-discovery with O(10~\invpb)
%\item Basic measurements calibrations with O(100~\invpb): \sigmattbar\ known to ~10-20\%, $\epsilon_{b}$ to 5 to 10\% 
%\item Precise measurements with O(1-10~\invfb):  JES to 1-2\%, \sigmat\ in the t-channel known to ~20\%, Top mass known to 1 to 3 GeV. 
%\item Exploration beyond SM with O(10~\invfb): looking for FCNC rare decays, small cross section resonances.

%\end{itemize}

%The analyses are currently being re-optimized to cope with the
%different S/B at \roots = 10 TeV, while eagerly
%awaiting for the LHC to start operations.
\section*{Acknowledgments}

The author would like to thank Martine Bosman, Gustaaf Brooijmans, Tim Christiansen, Richard Hawkings and John Parsons 
for useful discussions and comments. Additional useful discussions with J\"{o}rgen Sj\"{o}lin and Anne-Isabelle Etienvre 
are gratefully acknowledged.

\end{document}